\documentclass{PoS}

\title{Status of space-based gamma-ray astronomy}

\ShortTitle{Space based gamma-ray astronomy}

\author{Rolf B\"uhler\\
        DESY, Platanenallee 6, 15738 Zeuthen, Germany\\
        E-mail: \email{rolf.buehler@desy.de}}

\abstract{Gamma-ray observations give us a direct view into the most extreme environments of the universe. They help us to study astronomical particle accelerators as supernovae remnants, pulsars, active galaxies or gamma-ray bursts and help us to understand the propagation of cosmic rays through our Milky Way. This article summarizes the status of gamma-ray observations from space; it is the write-up of a rapporteur talk given at the 34$^{th}$ ICRC in The Hague, The Netherlands. The primary instrument used in the presented studies is the Large Area Telescope on-board the Fermi Spacecraft, which images the whole gamma-ray sky  at photon energies between 20 MeV and 2 TeV. The Fermi mission is currently in its 8$^{th}$ year of observations. This article will review many of the exciting discoveries made in this time, focusing on the most recent ones.}

\FullConference{The 34th International Cosmic Ray Conference,\\
		30 July- 6 August, 2015\\
		The Hague, The Netherlands}

\usepackage{graphicx}
\usepackage{amsmath}
\usepackage{wrapfig}

\begin{document}

\section{Introduction}

The study of cosmic rays has been entangled with many areas of physics since the very beginning (``cosmic rays'' will refer to relativistic charged particles in outer space, leptons and hadrons throughout this article). For instance,  cosmic-ray observations led to the discovery of pion mesons; one of the first reports of their discovery was at the first International Cosmic-Ray Conference (ICRC) in Krakow in 1947 \cite{Sekido2012}. Almost 70 years later, the field continued to build new ramifications, with dark matter searches, gamma-ray and neutrino astronomy being the latest additions into a field we call today Astroparticle physics. The ICRC continues to be an important forum to present and discuss advances in this research area. This article summarizes the status on space-based gamma-ray astronomy based on the contributions to the 34$^{th}$ edition of this conference held in The Hague\footnote{The slides of the rapporteur talk can be found under http://www.rolfbuehler.net/files/buehler\_spacegamma.pdf}. To make the report more complete, some results obtained in the past few years which were not shown at this conference were also included. Accompanying articles will give a status report of related topics in cosmic-ray research based on the contributions in this conference: Ground-based gamma-ray astronomy \cite{Lemoine-Goumard015}, Solar and heliospheric phenomena \cite{Heber011}, Neutrino Astronomy \cite{Ishihara013}, Indirect Dark Matter searches \cite{Cirelli014}, Cosmic rays: direct measurements \cite{Maestro016}, Cosmic rays: air showers from low to high energies \cite{Verzi015}.

The  Fermi  Gamma-ray  Space  Telescope  (\emph{Fermi}) is the most sensitive instrument observing the gamma-ray sky today\footnote{The Astro-rivelatore Gamma a Immagini Leggero (AGILE) satellite mission observing the sky at similar photon energies suffered a malfunction to the reaction wheel in November 2009, and has only been observing parts of the sky at reduced sensitivity since then. No contributions from the AGILE collaboration were submitted at the 34$^{th}$ ICRC.}.  The primary instrument on-board \emph{Fermi} is the Large Area Telescope (LAT) which images the sky between 20 MeV and 2 TeV. A second instrument, the Gamma-ray Burst Monitor  (GBM), complements  the  LAT  observations  of  transient sources at photon energies between 8 keV and 40 MeV. Every three hours the LAT images the full sky. New instrument response functions (``Pass 8'') were released in June 2015, which significantly improved the sensitivity of the detector\footnote{http://www.slac.stanford.edu/exp/glast/groups/canda/lat\_Performance.htm}. Depending on galactic latitude, the point source sensitivity at $\approx$ 1 GeV for 10 years of observations varies between $\approx4 \times 10^{-12}$ ergs cm$^{-2}$ s$^{-1}$ and $4 \times 10^{-13}$ ergs cm$^{-2}$ s$^{-1}$ (measured in 4 bin per decade). The angular resolution depends strongly on energy, it is $\approx 5^\circ$ at 100 MeV and decreases to $\approx 0.1^\circ$ above 30 GeV. The energy resolution also varies as a function of energy between 5\% and 20\%.

At the writing of this article \emph{Fermi} went into its 8$^{th}$ year of data taking. Many exciting discoveries were made in this time; this article will review some of them focusing on the most recent developments. It should be noted, that this article does not discuss results from the INTEGRAL satellite mission, which observes the sky from the X-ray to the soft gamma-ray band. For recent results from this mission the reader is referred to \cite{Bouchet896,Diehl2014,Ubertini2014}.  The article is structured as follows: section \ref{sec:diff} will discuss the diffuse gamma-ray emission from our galaxy, emphasizing on the Galactic Centre region and the ``Fermi-Bubbles''. In section \ref{sec:acc} different cosmic-ray accelerators will be discussed, as supernovae remnants, pulsars, gamma-ray bursts and active galactic nuclei. In section \ref{sec:multmes} the connection between gamma-rays and the recent astrophysical neutrino detections will be discussed. Finally, an outlook on proposed future gamma-ray missions will be given in section \ref{sec:future}.
\clearpage

\section{Diffuse galactic gamma-ray emission}
\label{sec:diff}

\begin{figure}[htp]
\centering
\includegraphics[width=0.96\textwidth]{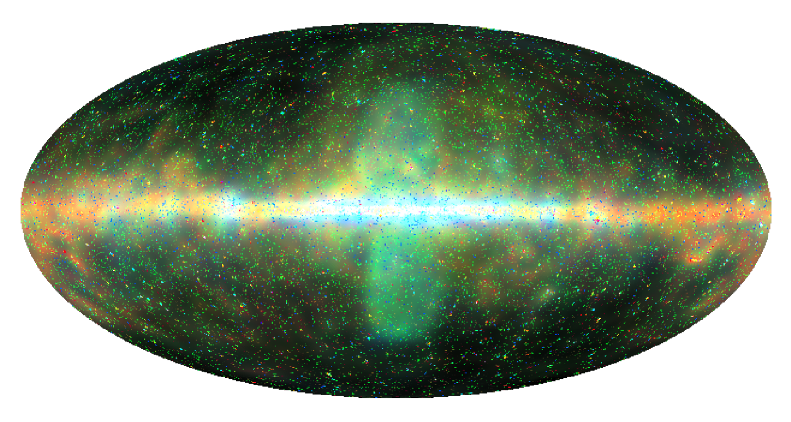}
\includegraphics[width=0.48\textwidth]{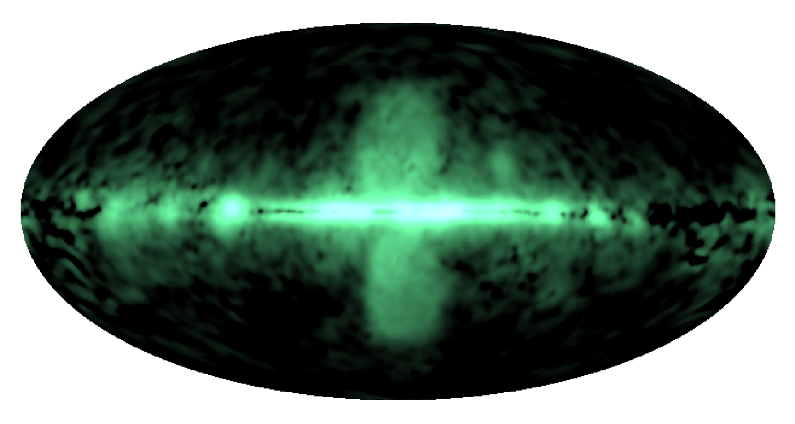}
\includegraphics[width=0.48\textwidth]{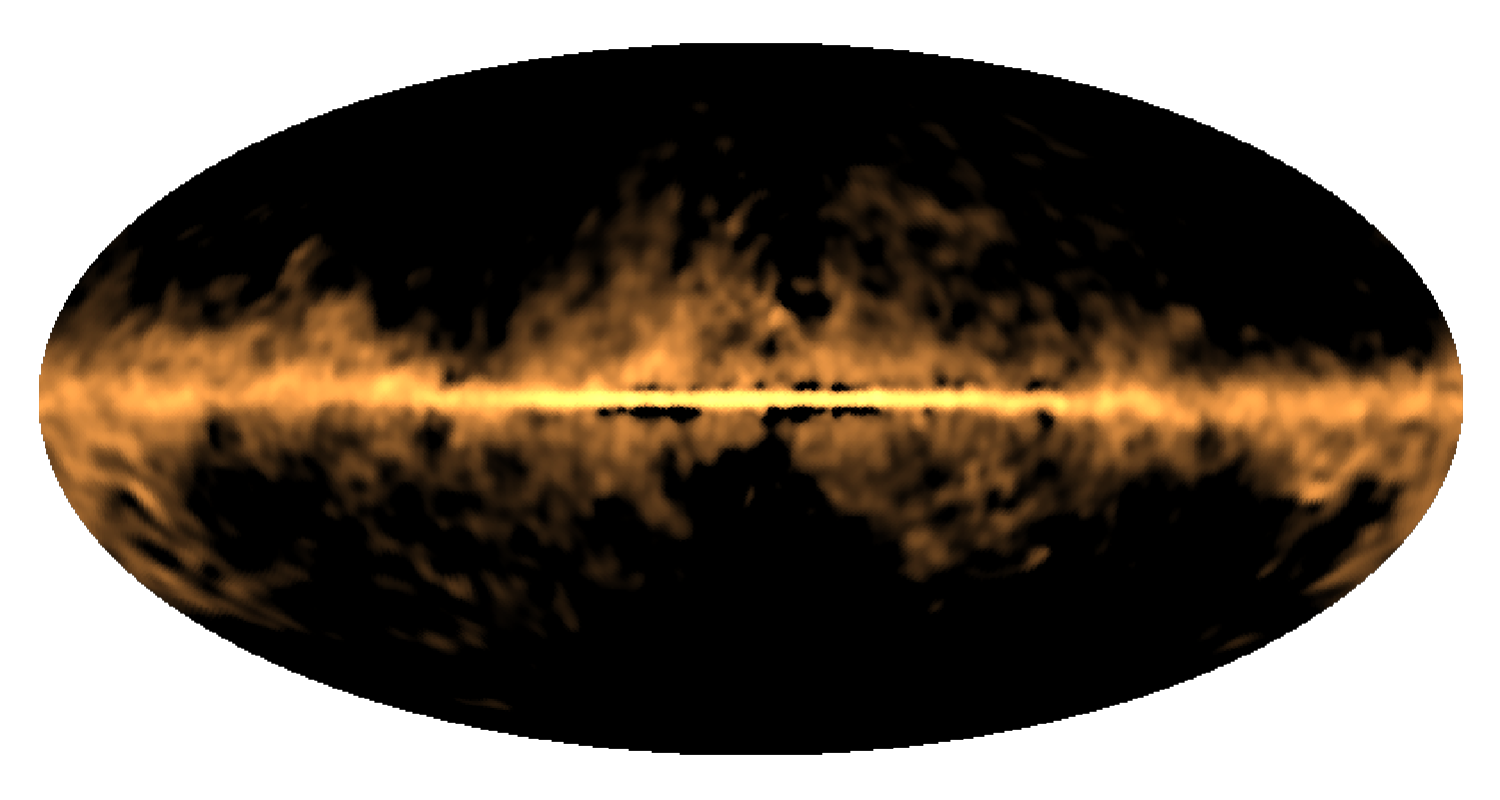}
\caption{The gamma-ray sky above 600 MeV in galactic coordinates. {\bf Top panel}: Diffuse galactic emission and point sources; {\bf bottom left panel}: Hard component of the galactic diffuse emission; {\bf bottom right panel}: Soft component of the galactic diffuse emission. The images have been de-noised and deconvolved using the D$^{3}$PO algorithm \cite{Selig2015}. The colour scale is designed to mimic the human perception of optical light in the gamma-ray range. Intensity indicates the (logarithmic) brightness of the flux, red corresponds to low-energy gamma rays around 1 GeV, and blue to gamma rays up to 300 GeV. The figures have been adapted from \cite{Selig768}.}
\label{fig:d3po}
\end{figure}

The gamma-ray sky seen by the LAT above 600 MeV is shown in the upper panel of figure \ref{fig:d3po}. The colour scale has been adjusted to mimic the human eye perception of optical light. Red colours show emission at photon energies of $\approx$ 1 GeV and green/blue emission photon energies of $\approx$ 100 GeV. The most prominent feature is the diffuse glow of our galaxy. Approximately 70\% of photons detected by the LAT from outside of our solar system are attributed to this emission. It is radiated by cosmic-ray protons, heavier nuclei and electrons interacting with interstellar gas and by cosmic-ray electrons interacting with photon fields via Inverse Compton (IC, throughout this article ``electrons'' refers to both, electrons and positrons).  Typically non-thermal distributions of particles in energy ($E$) -- protons ($p$), electrons ($e$) or gamma rays ($\gamma$) -- are well described by a power-law function $\frac{dN_{p,e,\gamma}}{dE} = N_0 E^{-\Gamma_{p,e,\gamma}} $, where $N_0$ is a normalization factor. In cosmic-ray interactions with interstellar gas on average a fraction of the energy of the primary is transferred to gamma rays: $E_\gamma \approx 0.1-0.3 E_{p,e}$). The spectral slope of the gamma-ray spectrum resulting from hadronic/Bremsstrahlung interactions will therefore follow the proton/electron one ($\Gamma_\gamma \approx \Gamma_p$, $\Gamma_\gamma \approx \Gamma_e$). In contrast, if electrons up-scatter photons via IC the resulting gamma-ray spectrum will be harder than the electron one ($\Gamma_\gamma = \frac{\Gamma_e + 1 }{2}$, in the ``Thomson regime'' which applies here in good approximation) \cite{Hinton2009}.

The cosmic rays responsible for the gamma-ray emission in the LAT energy range have energies between $\approx$ 1 GeV and $\lesssim$ 10 TeV. In this energy rage the local hadronic cosmic-ray spectrum falls off with a spectral index $\Gamma_p \approx 2.7$, and the electron spectrum with an index of $\Gamma_e \approx 3.3$.  If, in a first approximation  we assume the spectral slope of these particle distributions to be the same throughout the galaxy, we expect three spectral components in the galactic diffuse emission: (1) A soft component due to Bremsstrahlung emission of electrons with $\Gamma_\gamma \approx 3.3$ (2) A component due to hadronic cosmic-ray interactions with a gamma-ray spectral index of $\Gamma_\gamma \approx 2.7$ (3) A hard component due to IC scattering of electrons with a spectral index of $\Gamma_\gamma \approx 2.3$. The first two components are expected to closely follow the gas distribution in our galaxy, and therefore to be concentrated towards the galactic plane. Interestingly, this simple picture is qualitatively confirmed when the entire sky is decomposed into a soft ($\Gamma_\gamma =2.6$) and hard ($\Gamma_\gamma =2.4$) component \cite{Selig768,Selig2015}. The lower panels in figure \ref{fig:d3po} show the intensity attributed to each of these components in a fit in each pixel. The fit was done above 600 MeV, where the Bremsstrahlung contribution is thought to be less relevant. The soft component is therefore primarily expected to be due to hadronic interactions and indeed this component is concentrated on the galactic plane, qualitatively following the gas distribution. The hard component is presumably due to IC emission which extends to higher galactic latitudes; this is expected as the target photon fields decrease slower than the gas densities away from the galactic plane \cite{Porter908}. 

The main assumption made in the arguments above was that the cosmic-ray spectrum does not vary throughout our galaxy. Measurements of radioactive isotopes and spallation nuclei in the cosmic-ray flux reaching Earth suggest that particles are trapped for $\approx 10^7$ years inside the galactic disk at the energies of relevance here, after which they escape into the galactic halo \cite{Strong2007}. These measurements also show that the transport of cosmic rays through the galaxy is predominantly diffusive; in a first approximation the diffusion coefficient can be described as $D \approx 10^{28}~E^{0.45}$ GeV cm$^{2}$ s$^{-1}$. Cosmic rays in the energy range of interest here travel over a distance $\lesssim$3 kpc. Cosmic-ray electrons are confined to even smaller volumes due to radiation losses via synchrotron and IC emission. This means that in principle cosmic-ray densities can vary in different regions of the Milky Way. In addition, gas densities and photon fields depend strongly on the position in our galaxy. To take all of these effects into account self consistently is the aim of cosmic-ray propagation codes.

\subsection*{Cosmic-ray propagation codes}

Propagation codes solve the transport equations of cosmic rays numerically on a grid \cite{Strong2007}. The main input parameters are the neutral (HI), ionized (HII) and molecular H$_2$ gas distributions in our galaxy. Particularly the latter has large uncertainties, as it is usually only measured indirectly through Carbon monoxide lines. The models are constrained by the measured cosmic-ray abundances and their (an-)isotropy at Earth. As laid out in the previous section, direct cosmic-ray measurements at Earth only test our galactic neighbourhood, more distant regions are constrained primarily by the gamma-ray data.

\begin{wrapfigure}{r}{0.55\textwidth}
\centering
\includegraphics[width=0.53\textwidth]{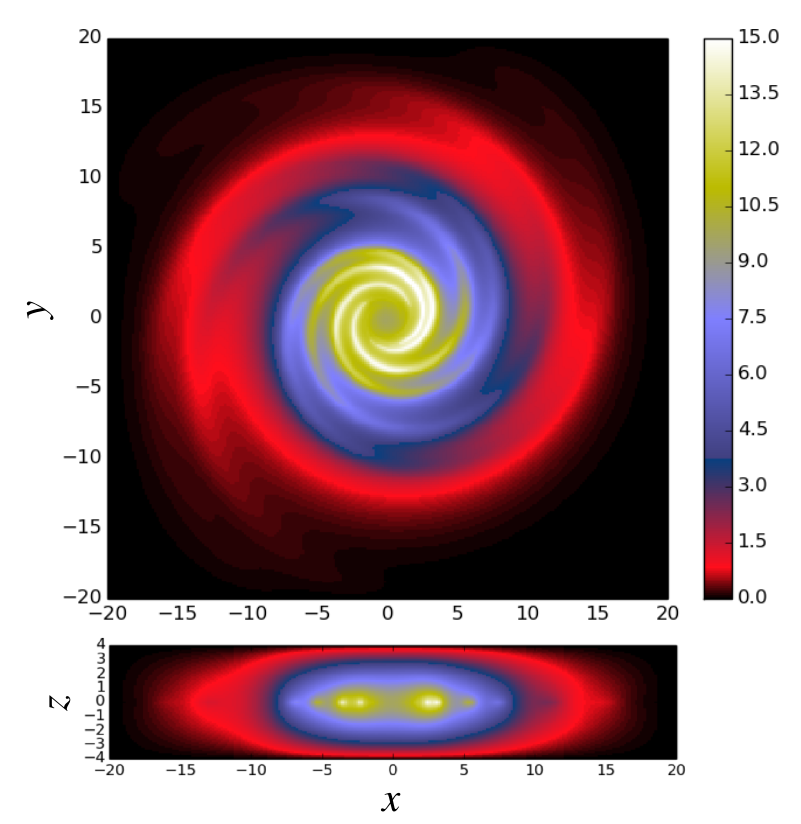}
\caption{Flux of $^{12}$C cosmic rays of energies $\approx$10 GeV in the Galactic plane seen from above the galactic plane (top) and from the side (bottom). Spatial scales are in kpc, the Earth is located at x = 8.5 kpc, y = z = 0 kpc. The figure is reproduced from \cite{Kissmann554}.}
\label{fig:picard}
\end{wrapfigure}
The most widely used propagation code is GALPROP \cite{Strong2007,Moskalenko492}. It is publicly available and can be run from a web interface\footnote{http://galprop.stanford.edu}. Over the recent years two new codes are being developed, PICARD \cite{Kissmann2014,Kissmann554} and DRAGON \cite{Evoli2008,Gaggero020}; the latter is also publicly available\footnote{http://www.dragonproject.org/Home.html}.  The aim of these codes is to make propagation models increasingly realistic; for instance, by taking into account the spiral-arm structure of gas in the Milky Way. This requires running the code in three dimensions (see figure \ref{fig:picard}), in contrast to axisymmetric two-dimensional models used till today. It was shown that the spiral-arm structure has a strong effect on the predicted gamma-ray fluxes of up to 50\% \cite{Johannesson517}. Other recent improvements are the inclusion of direction dependent diffusion coefficients. An important ingredient is also the infra red and optical photon distribution in our galaxy, which affects the expected inverse Compton emission. These photon densities are difficult to measure directly due to foregrounds and therefore model predictions are used, which keep increasing in complexity and realism \cite{Porter908}.

\begin{figure}[t]
\centering
\includegraphics[width=0.98\textwidth]{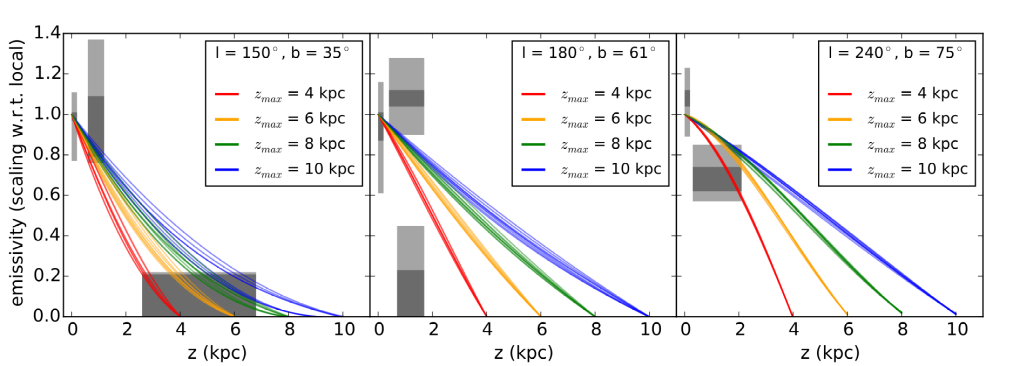}
\caption{ Gamma-ray emissivities (gamma-ray flux per target mass) for different High and Intermediate Velocity Clouds, as a function of the scale height $z$ above the galactic disk. The emissivities have been normalized to 1 in our galactic neighbourhood. Dark grey boxes show statistical uncertainties, light grey boxes systematic ones. The coloured lines are expected emissivity curves from cosmic-ray propagation models of different halo sizes \cite{Tibaldo2015}. The figure is reproduced from \cite{Tibaldo751}.}
\label{fig:hvc}
\end{figure}

An important parameter in propagation models is the height above the galactic disk where cosmic rays escape from our galaxy (the ``scale height'' of the galactic halo, see lower panel of fig. \ref{fig:picard}). This quantity is highly uncertain and could till now only be inferred indirectly through measured cosmic-ray abundances at Earth. However, the first direct measurements were done recently by measuring the gamma-ray flux from High and Intermediate Velocity Clouds \cite{Tibaldo2015,Tibaldo751}. These clouds are thought to be gas falling onto the Milky Way, as they do not participate in the galactic rotation. For some of them the distance can be measured, primarily by observing absorption patterns in stars located behind and before them. As can be seen in fig. \ref{fig:hvc}, these measurements indeed show a decrease of the cosmic-ray density above the galactic disk. These measurements are expected to improve in accuracy as the distance to the clouds is better determined, for instance with the help of the GAIA satellite which is currently taking data.

\subsection*{The Galactic Centre}

A region of particular interest in our Galaxy is the Galactic Centre (GC) region. It contains the super massive black hole Sgr A$^*$; the inner $\approx$100 pc around Sgr A$^*$ are thought to be the birth place of $\approx$  1 \% of the stars in our Milky Way. Cosmic-ray propagation models have particularly large difficulties in this region for several reasons: (1) The distance measurements of gas clouds are uncertain, as rotation curves diverge at the GC (2) The environment is very turbulent, advection is expected to play a larger role than on average in the galaxy (3) Advection and diffusion is expected to be anisotropic. In addition, it is observationally difficult to constrain the cosmic-ray densities close to the GC. Gamma-ray measurements are dominated by large foreground and background emission in the line of sight through our Milky Way; only 5--10 \% of the gamma-ray emission coming from the direction of the GC is expected to come from the inner 1 kpc \cite{Calore2015}.

\begin{figure}[tp]
\centering
\includegraphics[width=0.98\textwidth]{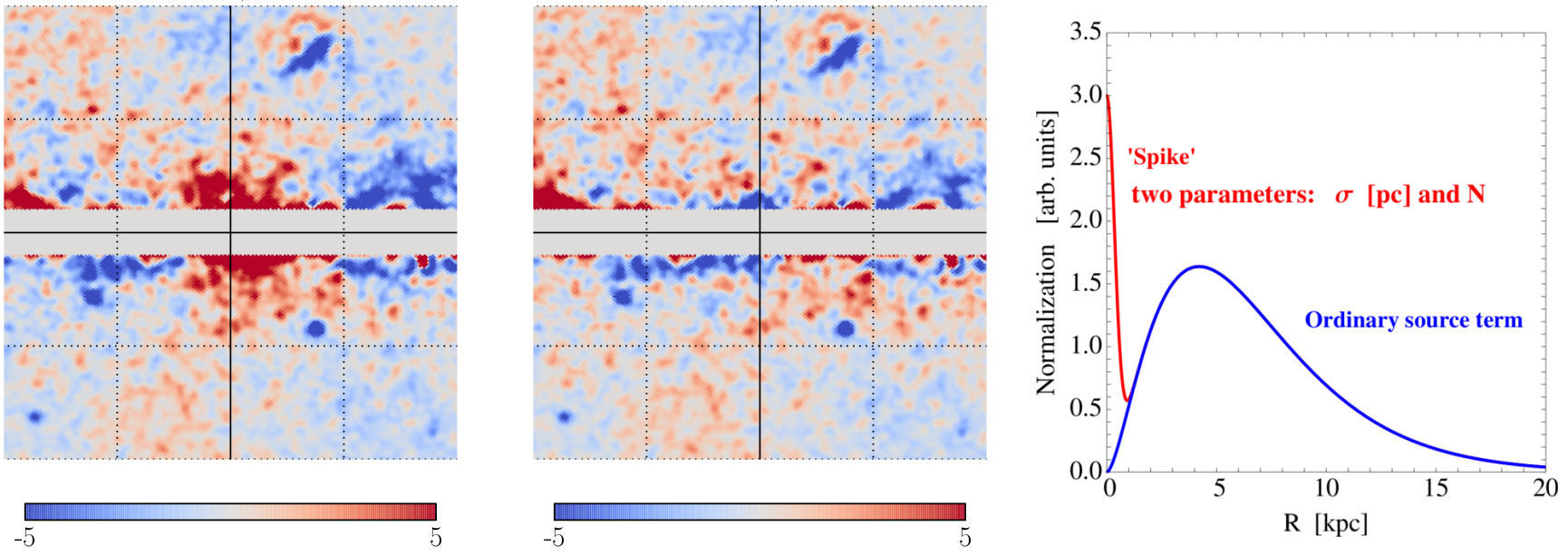}
\caption{Residual counts maps between the prediction of a cosmic-ray propagation model and the observations in the galactic centre region at photon energies of 1$-$10 GeV \cite{Gaggero2015,Gaggero020}. The {\bf right panel} shows the cosmic-ray source distribution assumed in the propagation model. The typically assumed source distribution is shown by the blue line. A source term accounting for the high star formation rate in the galactic centre region is shown by the red line. The {\bf middle panel} shows the residual map if both source terms are included. The {\bf left panel} shows the residual map if the additional sources in the galactic centre region are omitted in the propagation code. The figure has been adapted from \cite{Urbano909}.}
\label{fig:gcspike}
\end{figure}

When comparing the predicted gamma-ray flux from cosmic-ray propagation codes in the GC region with measurements by the LAT, several authors have pointed out that there is an excess of emission around 1 GeV. This excess was found to be extended with consistent spectral characteristics within $\approx 15^\circ$ around the GC \cite{Calore2015}. Several explanations for the excess have been proposed, as an unresolved population of millisecond pulsars \cite{Weniger920}, recurring outbursts from Sgr A$^*$ \cite{Calore915} and Dark Matter annihilation \cite{Cirelli014}. However, given the uncertainties mentioned before, our current knowledge of the cosmic-ray distribution in the galactic centre is not good enough to robustly confirm or refute these ideas. For instance, the models used to detect the gamma-ray excess are based on a source distribution of cosmic rays which follows the pulsar distributions. When taking into account the increased star formation rate expected in the GC the excess largely goes away, as shown in fig. \ref{fig:gcspike}. It is also likely that an increased infra red photon density in the GC region leads to a significantly higher gamma-ray emission via IC than expected in the models used to measure the gamma-ray excess \cite{Porter815,Tuffs923}.

\subsection*{The Fermi Bubbles}

One of the most prominent features visible in fig. \ref{fig:d3po} are the ``Fermi Bubbles''. They extend out to $\approx 50^\circ$ north and south of the GC. The spectra of the northern and southern bubbles are found to be the same within the measurement uncertainties: the spectrum is hard with a spectral index $\Gamma_\gamma \approx 1.9$, which turns down above energies of $\approx$ 100 GeV \cite{Su2010,Ackermann2014} . If the bubbles are located at the distance of the GC (as their symmetry suggests) their size is $\approx$ 10 kpc. The energy in cosmic rays inside of them to explain the gamma ray emission is $\approx 10^{52}$ ergs in the case of a leptonic origin of the emission and  $\approx 10^{55}$ ergs for an hadronic origin of the emission \cite{Ackermann2014}. 

The hard spectrum of the bubbles points towards a leptonic origin due to IC scattering (Brems- strahlung emission is expected to be less relevant here, as the gas densities are expected to be low $n_H \lesssim 0.01$ cm$^{-3}$). In this scenario, synchrotron emission of the same electrons can naturally explain a similar feature observed at micro waves (the ``WMAP haze'' \cite{Hooper2007}). Comparing the ratio of the synchrotron to IC emission gives an estimate of the magnetic field inside the bubbles of $\approx 8 \mu$G \cite{Ackermann2014} . The main difficulty with leptonic scenario is that electrons of the required energies ($\approx 1$ TeV) are expected to cool before reaching the upper and lower edge of the bubbles. However, this can be overcome in the case that there is reacceleration due to turbulence inside the bubble outflow \cite{Mertsch2011}, or if the electron transport is dominated by fast advection (with speed $\gtrsim$ 10~000 km s$^{-1}$) \cite{Ackermann2014}. The latter implies a burst like emission of plasma, which might be produced due to past activity of Sgr A$^*$ \cite{Guo2011}. Alternatively, the bubbles can also be explained by hadronic scenarios. The much longer cooling time of hadrons allows for a steady injection of these particles into the bubbles over long time scales. This might for instance be due to the stellar winds induced by the high star formation rate near the GC. However, in this scenario secondary electrons produced in the hadronic interactions fail to describe the spectrum of the WMAP haze emission; additional injection of cosmic-ray electrons is therefore still necessary \cite{Crocker906}.

\clearpage 
\section{Cosmic accelerators}
\label{sec:acc}

\begin{wrapfigure}{r}{0.55\textwidth}
\centering
\includegraphics[width=0.53\textwidth]{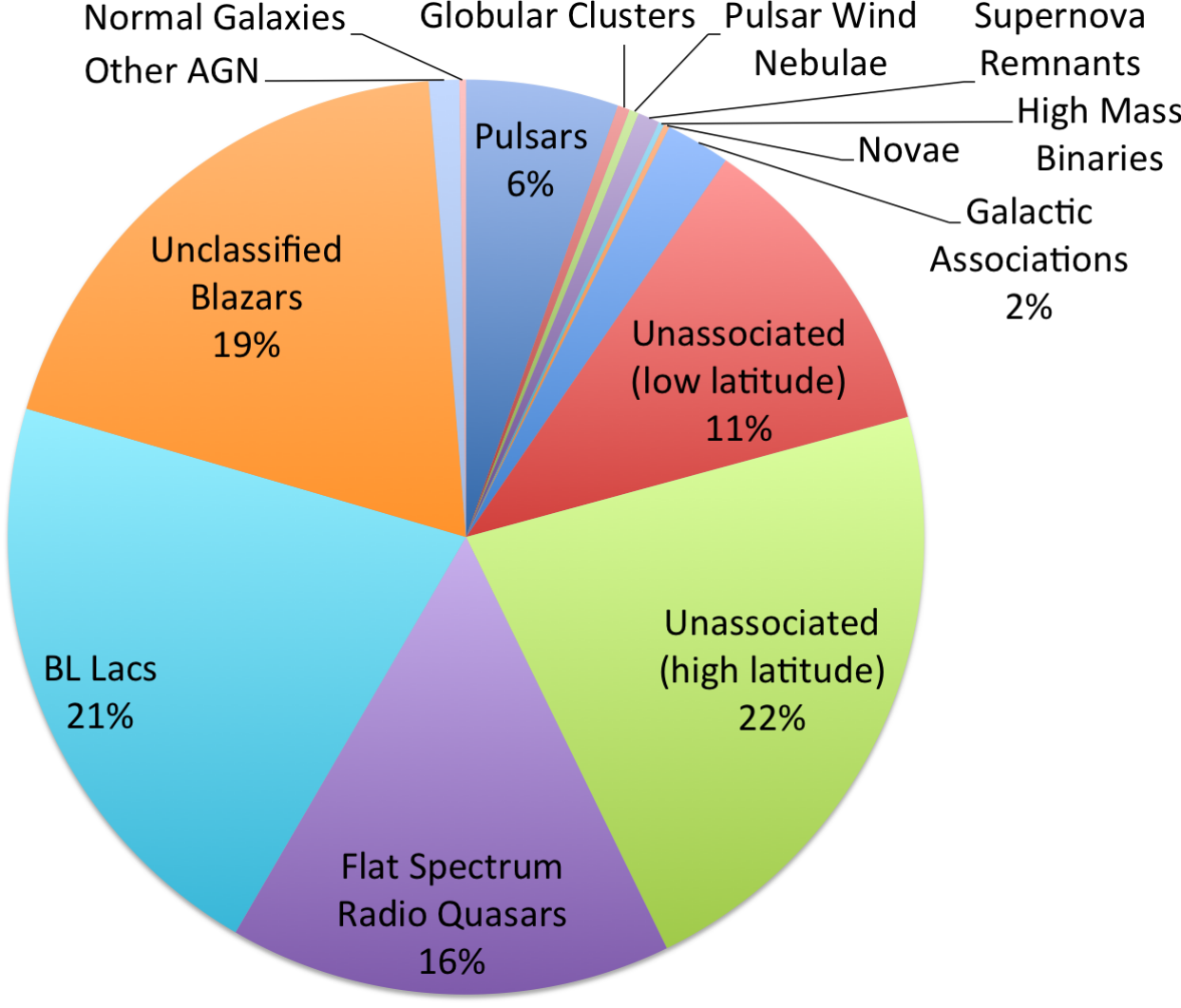}
\caption{Number of sources detected for different source classes in the 3FGL catalogue\cite{Acero2015}. The figure is reproduced from \cite{Hays010}.}
\label{fig:3fglclass}
\end{wrapfigure}

\begin{table}[bp]
\centering
\begin{tabular}{llccc}
    Catalogue & Description & Time [months] & Energy [GeV] & Reference\\ \hline
	3FGL & Standard source catalogue & 48 & 0.1 - 100 & \cite{Acero2015,Ballet848} \\
	2FHL & Hard sources catalogue & 80 & 50 - 2000 & \cite{Ajello019,Ackermann2015b} \\
	2PC & Pulsar catalogue & 36 & 0.1 - 100 & \cite{Abdo2013} \\
	3LAC & Active galactic nuclei catalogue & 48 & 0.1 - 100 & \cite{Gasparrini879} \\
	1FAV & Flaring sources list & 47 & 0.1 - 300 & \cite{Ackermann2013} \\
    1DF  & Point-source candidate list& 78 & 0.6 - 300 & \cite{Selig2015,Selig768} \\
    GBM-GRB & Gamma-ray bursts by the GBM & 48 & 10$^{-5}$-0.01 & \cite{Kienlin2014,Gruber2014} \\
	LAT-GRB & Gamma-ray bursts by the LAT & 36 & 0.02 - 100 & \cite{Ackermann2013b,Bissaldi796} \\
	1SNR & Supernovae remnants & 42 & 1.0 - 100 & \cite{Brandt722,dePalma814} \\
\end{tabular}
\caption{List of source catalogues based on \emph{Fermi} data.}
\label{tab:cats}
\end{table}

The objects accelerating cosmic rays appear as point sources in the gamma-ray sky (see fig. \ref{fig:d3po}). Particle acceleration is common in the universe, as many sources have been found to accelerate cosmic rays on very different spatial scales; from jets of AGN on scales of $\sim$ kpc, to pulsars with sizes of $3 \times 10^{-13}$ pc ($\approx 10$ km). With more than 3000 sources detected to date, the LAT has moved into ``an era of catalogues''\footnote{This term was used by Liz Hays in her overview talk at the conference \cite{Hays010}.}. Source properties are quantified in uniform ways and studied for source type populations.  A list of the available catalogues using LAT data are shown in table \ref{tab:cats}. The different types of sources detected in the primary LAT catalog (3FGL) are shown in fig. \ref{fig:3fglclass}. Most of them are ($\approx$ 80 \%) are extragalactic, with blazars (AGN where the jet is pointing towards the Earth) making up the majority of the sources. Among the galactic sources pulsars are the most common gamma-ray emitters. A brief discussion of the individual source classes will be given later in this section.

\begin{figure}[tp]
\centering
\includegraphics[width=0.98\textwidth]{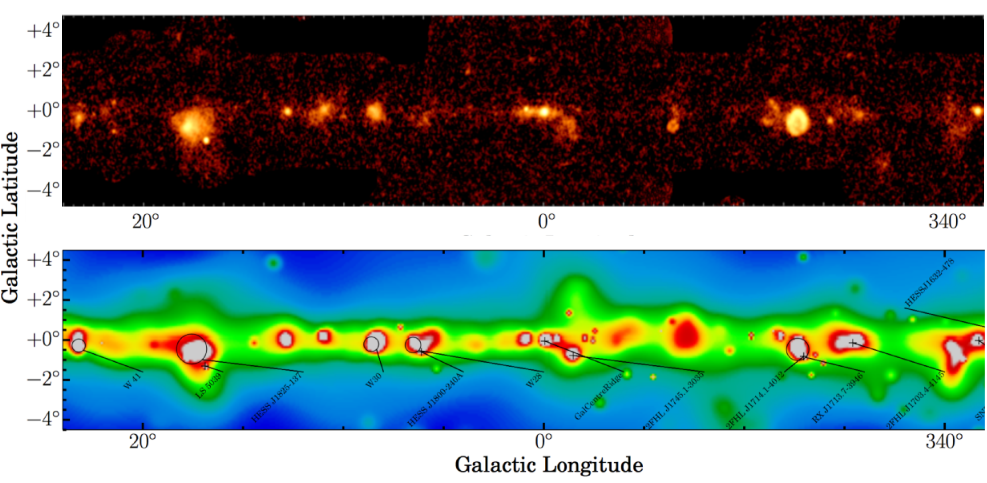}
\caption{Gamma-ray view of the inner Milky way in gamma rays. {\bf Upper panel}: Significance map of the H.E.S.S. galactic plane scan at photon energies above 200 GeV with a point source sensitivity of $\approx$1\% of the Crab nebula flux (preliminary results) \cite{Deil773, Aharonian2006}. {\bf Lower panel}: Smoothed counts map t photon energies 50 GeV measured by the \emph{Fermi}-LAT with a point source sensitivity of $\approx4$\% of the Crab nebula flux. The figure has been adapted from \cite{Ajello019}.}
\label{fig:2fhlhess}
\end{figure}

The first catalogue using the improved instrument responses of the LAT was presented at the 34$^{th}$ ICRC for the first time, the ``hard sources catalogue'' (2FHL) \cite{Ackermann2015b}. It searches for sources emitting above 50 GeV. In total 360 sources were detected above this energy, out of which 57 were previously unknown gamma-ray emitters. Particularly interesting is the comparison to the scan of the inner Milky Way performed by the H.E.S.S. telescopes above 200 GeV, due to the overlap in energy range. A direct comparison of the source positions is shown in figure \ref{fig:2fhlhess}. Five previously unknown extended sources are detected in the 2FHL. While the H.E.S.S. source population is dominated by pulsar wind nebulae (PWN), Supernovae Remnants (SNR) are the dominant source class among the galactic 2FHL sources. 

\subsection*{Supernovae Remnants}

Supernovae have long been suspected to be the primary sources of cosmic rays up to energies of $\approx$ 1 PeV. One of the most important arguments is that they are one of the few source classes which can provide the required energy to sustain the cosmic-ray flux if $\approx$10 \% of the energy released in each explosion is channelled into accelerated particles. That SNRs are gamma-ray emitters, shows that they accelerate particles to energies beyond 1 TeV. However, whether these particles are leptons or hadrons remains unclear. 

\begin{wrapfigure}{r}{0.6\textwidth}
\centering
\includegraphics[width=0.58\textwidth]{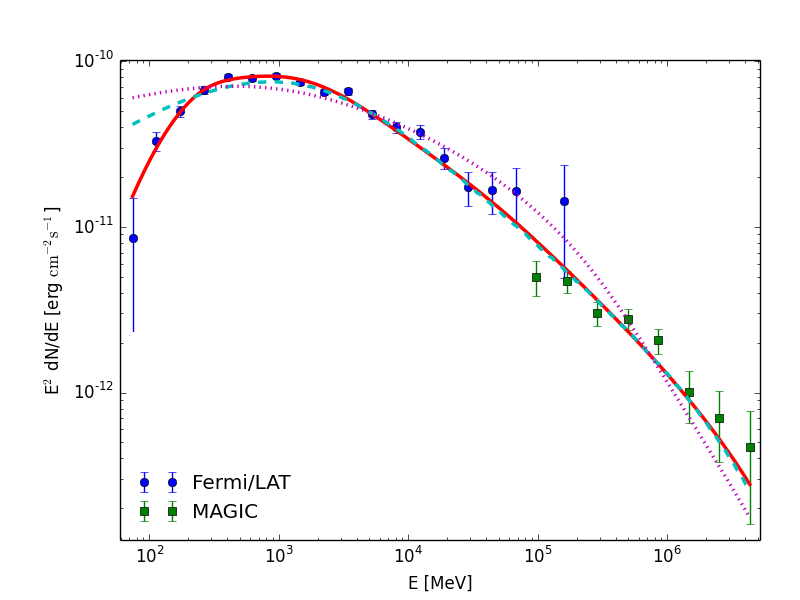}
\caption{Gamma-ray spectrum of the supernova remnant W51C (preliminary results). The ``pion bump'' is visible at the low energy end, indicating a hadronic origin of the emission. A hadronic emission model is shown by the red straight line. Two leptonic models are also shown (Inverse Compton purple dotted line, Bremsstrahlung magenta dashed line).  The figure has been adapted from \cite{Jogler888}.}
\label{fig:w51c}
\end{wrapfigure}

Over the past years the population of gamma-ray emitting SNRs is increasing, with more than 30 of them detected in the energy range of the LAT \cite{Brandt722,dePalma814}. The picture which is emerging from this sample is that whether the gamma-ray emission is produced by hadrons or leptons depends strongly on the environment and age of the system: hadronic emission dominates if the SNR is interacting with molecular clouds, while leptonic emission dominates for non interacting young SNR \cite{Brandt722,dePalma814}. Two examples for this are the SNRs RCW86 and W51C. The first one is a young SNR, likely linked to the historical supernova SN~185. The gamma-ray spectrum is very hard ($\Gamma_{\gamma,p} < 1.7$) \cite{Brun866}. As simulations of SNR shocks show that spectra obtained in SNRs are typically not harder than $\Gamma_{p} \simeq 2$ \cite{Caprioli008,Pohl455}, this strongly suggests a leptonic origin of the gamma-ray emission. In contrast, W51C is a middle aged SNR with an age of $\approx$~30~000 years, interacting with molecular clouds. As shown in fig. \ref{fig:w51c}, the gamma-ray spectrum in this system shows the ``pion bump'' expected for a hadronic origin of the emission \cite{Ackermann2013c,Jogler888}.

\subsection*{Binary systems}

To date, a handful of high-mass X-ray binaries systems have been detected at gamma rays \cite{Dubus2013}. In these systems the emission is modulated with the rotation period of a compact object around a massive star. It remains unclear for most of them whether the compact object is a neutron star or a black hole.  Recently two new classes of binary systems where discovered at gamma rays: novae and a colliding wind binary, Eta Carinae. In the latter two massive stars -- an O or WR star with M $\approx$ 30 M$_\odot$ and a luminous blue variable star with M $\approx$ 120 M$_\odot$ -- orbit each other with a period of 5.54 years. Recently, the first full orbit of Eta Carinae was measured by the LAT \cite{Reitberger2015}. Interestingly, a new spectral component is detected above $\approx$ 10 GeV during the periastron passage (when the stars are closest to each other); this can be interpreted as acceleration of hadrons during this period. Generally, it appears difficult to accelerate electrons to the required energies in these systems as radiation losses are strong due to IC in the intense stellar photon fields \cite{Ohm913,Ohm2015}. However, whether or not leptonic emission can still dominate at some orbital phases remains unclear \cite{Reimer921}.

In novae explosions, a white dwarf (WD) and a star orbit each other. As the WD accretes material from its companion, a thermonuclear explosion is ignited at its surface, producing the nova explosion. The first detection of such systems was a symbiotic binary system, where the companion is a red giant star. However, more recently the LAT has detected 5 further novae, where the companion is a main sequence star \cite{Cheung880,Ackermann2014c}. In these so called ``classical novae''  the explosion is expected to occur into a much weaker stellar wind. It is therefore surprising that the characteristics of all six novae in gamma rays are very similar: the flare lasts for $\approx$ 2 weeks and the energy spectrum is curved, falling of sharply above a few GeV. Classical novae are expected to happen 20--50 times per year in our Milky Way. All the novae detected by the LAT were within a distance of $\approx$ 5 kpc. It is therefore possible that all novae produce gamma rays and the detectability is primarily determined by the distance of the explosion and the instrument sensitivity of the LAT.

\subsection*{Pulsars}

Pulsar physics has made large steps forward since the beginning of LAT observations. The population has grown from only six gamma-ray pulsars known before the launch of Fermi to over 160 today \cite{Caraveo2014}. As it often happens in science, these new insights have led to many new questions: the hope to explain the difference between pulsars in one model, varying only the observing angle and the angle of the magnetic moment of the neutron star towards its rotation axis, has not been successful so far. Instead, the gamma-ray emission might originate from (several) different regions in the magnetosphere for different pulsars (see for example \cite{Pierbattista2015}). The magnetospheric structure can even undergo sudden changes for single systems, as shown by the recent discovery of a pulsar with rapid changes in spin-down frequency ($\dot P$) and gamma-ray flux \cite{Allafort2013}. The study of PWN has also revealed much more complexity in the outflows of pulsars than was previously expected: strong gamma-ray outbursts were discovered from the Crab nebula. These events are likely the result of magnetic reconnection in the relativistic wind emitted by the pulsar. How close to the magnetosphere this happens remains uncertain \cite{Buehler2014,Rudy2015}.

Almost 50\% of the known gamma-ray pulsars detected to date belong to the class of millisecond pulsars (MSPs, with periods $P \lesssim $ 30 ms) \cite{Renault843,Laffon881}. These systems are clear outliers in the $P - \dot P$ plane \cite{Caraveo2014}. They are thought to be spun-up to high rotation frequencies by accreting matter from companion stars. Indeed, several systems have been found in the few past years which undergo changes from rotation power states, to accretion powered ones \cite{Bednarek892}. These findings indicate that MSPs are intimately connected to low mass X-ray binary systems. The first gamma-ray emitting pulsar for which such a state transition was observed is PSR J1023+0038 \cite{Stappers2014}. 

\subsection*{Active Galactic Nuclei}

\begin{figure}[tp]
\centering
\includegraphics[width=0.98\textwidth]{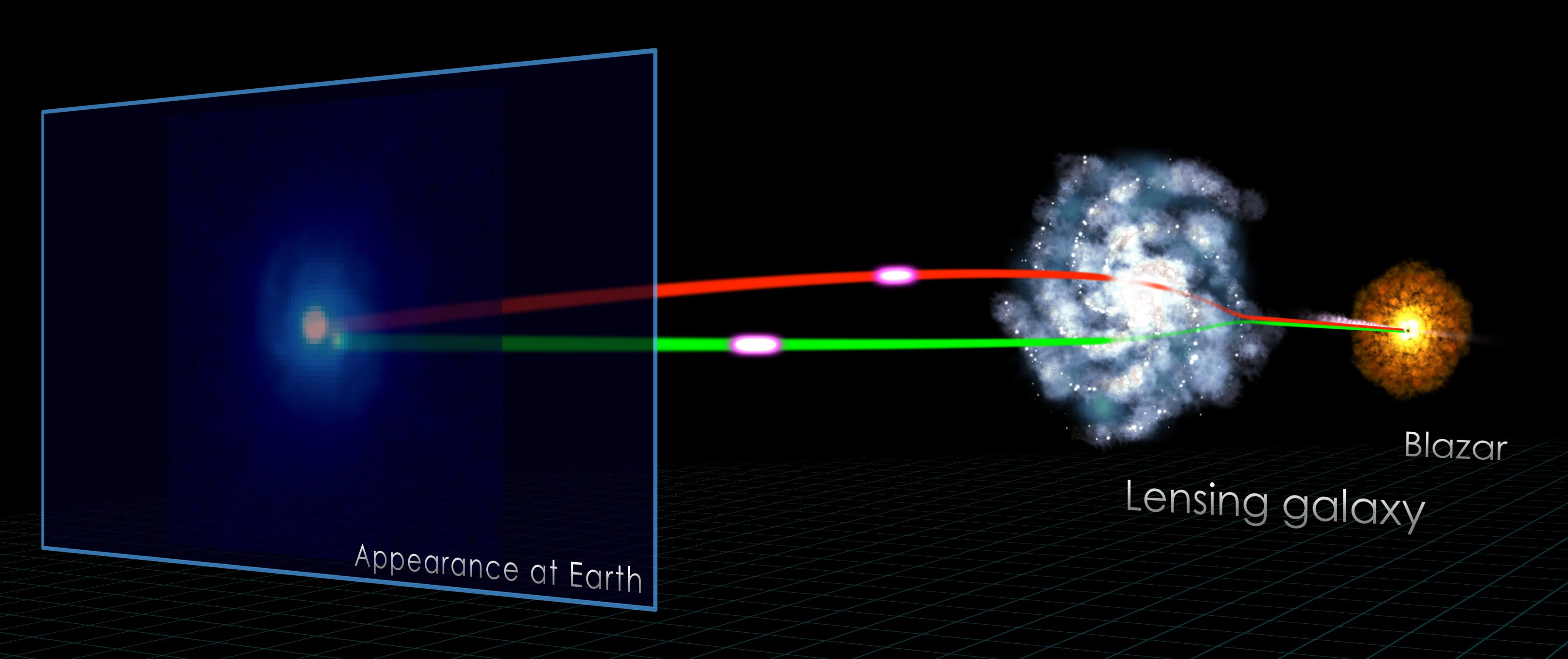}
\includegraphics[width=0.98\textwidth]{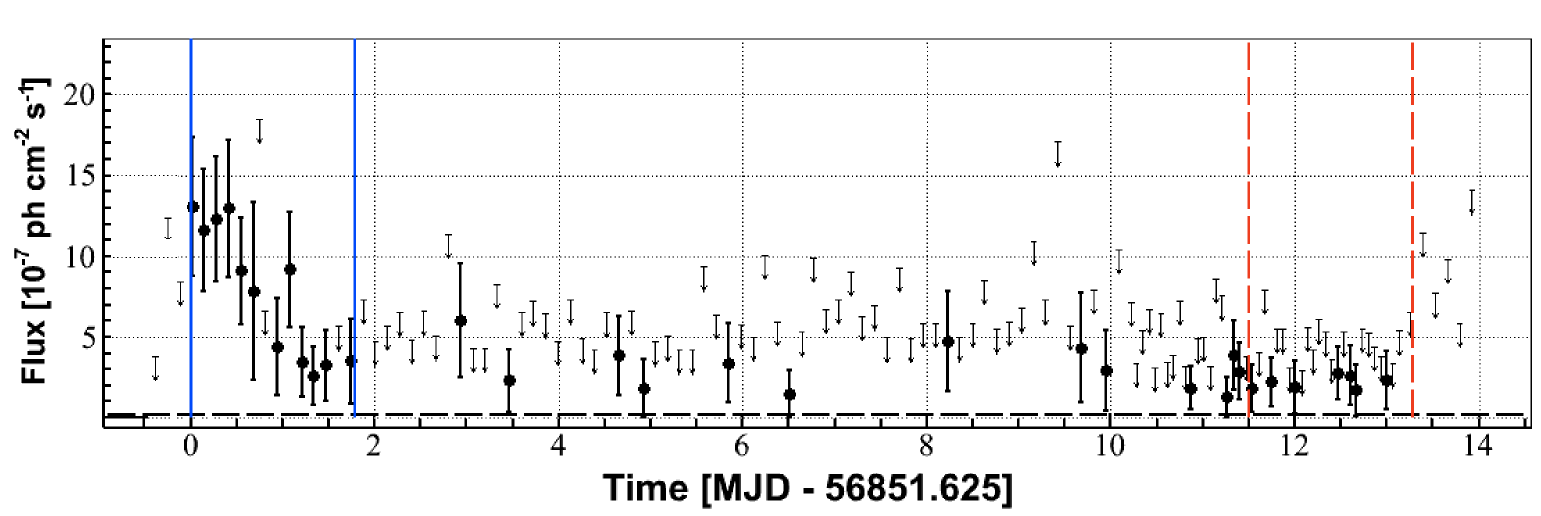}
\caption{The gravitationally lensed blazar B0218+357 shown in a schematic view (top panel, image credit: NASA's Goddard Space Flight Center) and its gamma-ray flux above 100 MeV as a function of time (lower panel, preliminary results). The blue vertical bars indicate the time window of a flare detected by the \emph{Fermi}-LAT in 2014. According to the time delay expected due to the gravitational lensing measured during 2012, a second flare was expected 11.46 $\pm$ 0.16 days later, indicated by the vertical dashed red lines \cite{Cheung2014}. The second flare, if present, was much weaker than expected. The bottom figure has been adapted from \cite{Buson877}.}
\label{fig:agnlensed}
\end{figure}

The extragalactic gamma-ray sky seen by the LAT has recently been reviewed \cite{Massaro2015}; here only some findings will be highlighted. With more than 1500 sources, AGN are the largest source population in the gamma-ray sky \cite{Gasparrini879,Tomono717,Reimer910}. A redshift dependent absorption feature in a sample of AGN was detected, which emerges due to gamma-ray absorption by the Extragalactic Background Light (EBL). This feature was used to measure the intensity of the EBL between optical and the ultra violet frequencies \cite{Ackermann2012,Dominguez718}. Another finding is that optical polarization changes in blazars are correlated with the emission of gamma-ray flares, revealing highly aligned magnetic fields in the gamma-ray emitting regions \cite{Abdo2010}.

Recently, the first two gravitationally lensed blazars have been detected by the LAT: PKS 1830-211 \cite{Neronov2015} and B0218+357 \cite{Buson877,Vovk2015}. In these systems two (or more) images are obtained from one object in the optical and radio. Due to the limited angular resolution, these two images cannot be resolved by the LAT. However, due to the different travelling times around the lensing galaxy, a time delay of the signal can be detected. For B0218+357, a time delay of 11.46~$\pm$~0.16 days was detected for the flaring activity of 2012 \cite{Cheung2014}. In 2014, a new flare was detected, as shown in fig. \ref{fig:agnlensed}. However, the expected delayed flare (if present) was much weaker than expected. This points toward changes in the micro lensing properties in the lensing galaxy. If this interpretation is correct, it sets stringent limits on the size of the gamma-ray emission region $R_\gamma$, as micro lensing can only be active if the latter is smaller than the Einstein radius $R_E$ of the micro-lensing stars $R_\gamma \lesssim R_E \approx 3 \times 10^{16}$ cm \cite{Vovk2015}.

\subsection*{Gamma-ray Bursts}

The GBM detects $\approx 20$ bursts per month at photon energies between 10 keV to 40 MeV; out of these $\approx 1$ burst per month is detected at higher energies by the LAT \cite{Bissaldi796,Kienlin2014,Gruber2014,Ackermann2013b}. This large sample of GRBs shows a wide variation in burst characteristics (there is a saying in the GRB community ``If you have seen one GRB, you have seen one GRB''). The observed energy spectra were generally more complex than known before the \emph{Fermi} mission. The phenomenological Band function is not a good representation of the spectrum for many bursts, new spectral components were sometimes found to emerge at the high-energy end. A common characteristic is that the energy fluence in the GBM energy range is larger than the one in the LAT and that the LAT emission is delayed in time with respect to emission at lower frequencies. Also, the LAT emission appears to be more extended in time. 

One of the most energetic bursts detected so far was GRB 130427A (labelled ``The Monster'') \cite{Bissaldi796}. As it also happened to occur nearby ($z = 0.34$) it gave us the most detailed view into the time development of the gamma-ray emission; the highest-energy photon (95 GeV) and the longest duration (20 hours) were detected during this burst. In the standard model of GRBs, a blast wave from an imploding massive star produces the initial prompt emission (alternatively, the blast wave can be emitted during neutron star and/or black hole mergers). The blast wave collides with the external surrounding material and creates shocks. These external shocks accelerate charged particles, which produce photons through synchrotron radiation. The emission detected by the LAT is thought to be emitted in this second phase, providing a natural explanation for the observed time delay. However, the high-energy of the gamma-ray emission observed during GRB 130427A is difficult to explain in this picture, as strong synchrotron losses are expected to limit the energy of the accelerated particles \cite{Ackermann2014b}.

\section{Multi-messenger Astronomy}
\label{sec:multmes}

\begin{wrapfigure}{r}{0.55\textwidth}
\centering
\includegraphics[width=0.53\textwidth]{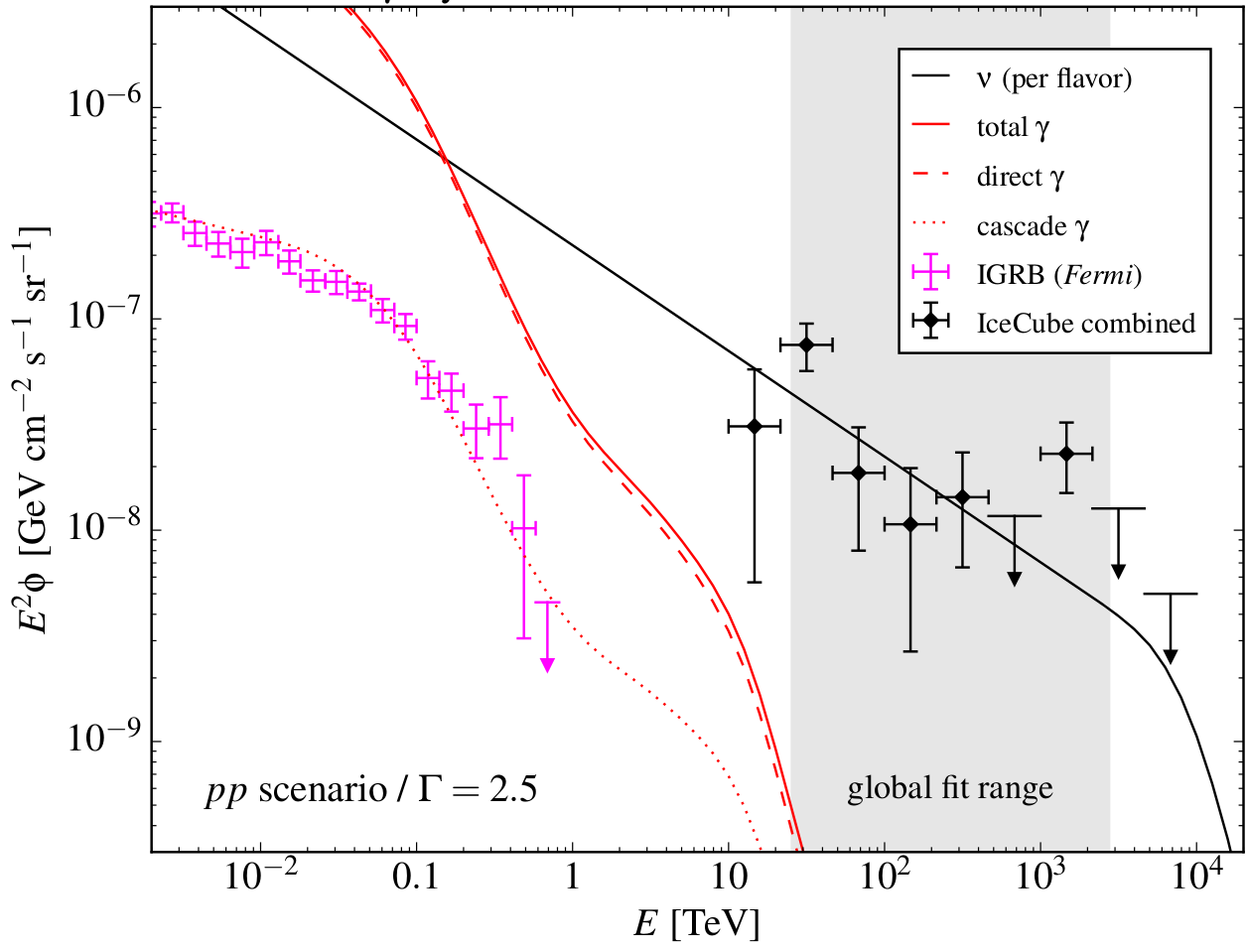}
\caption{The Isotropic Gamma-Ray Background detected by the \emph{Fermi}-LAT \cite{Ackermann2015} and the diffuse neutrino emission detected by Ice-Cube \cite{Aartsen2015}. The figure is reproduced from \cite{Ahlers022}.}
\label{fig:multmes}
\end{wrapfigure}

The window to multi-messenger astronomy has  been opened at high energies with the first detection of astrophysical neutrinos by IceCube \cite{Ishihara013}. No point sources were found yet, the observed neutrino flux is compatible with being isotropic. The energy spectrum is well described by a power law with a spectral index of $\Gamma_{\gamma} = 2.50 \pm 0.09$ \cite{Aartsen2015}. If the neutrinos are produced in p--p collisions of cosmic rays with gas, gamma rays are also produced in these interactions. It is therefore interesting to compare the neutrino emission to the Isotropic Gamma-Ray Background (IGRB) observed by the LAT. The IGRB accounts for $\approx$ 50 \% of the extragalactic gamma-ray flux (the other $\approx$~50~\% are due to resolved point sources) \cite{Ackermann2015}. The comparison is shown in fig. \ref{fig:multmes}. An interesting conclusion can be drawn: if the neutrinos are indeed created in p--p interactions, there has to be a spectral break in the proton spectrum below the IceCube energy range; otherwise the gamma-ray flux would be almost one order of magnitude higher than the IGRB \cite{Ahlers022}. However, other possibilities are that the gamma rays are absorbed inside of the source, or that the neutrinos are produced in cosmic-ray interactions with photons. While these caveats do not allow such studies to be conclusive yet, they show the power of the  multi-messenger approach, which is going to become increasingly important once neutrino sources are found and measurements become more accurate.  

\section{Future Missions}
\label{sec:future}

\begin{wrapfigure}{r}{0.55\textwidth}
\centering
\includegraphics[width=0.53\textwidth]{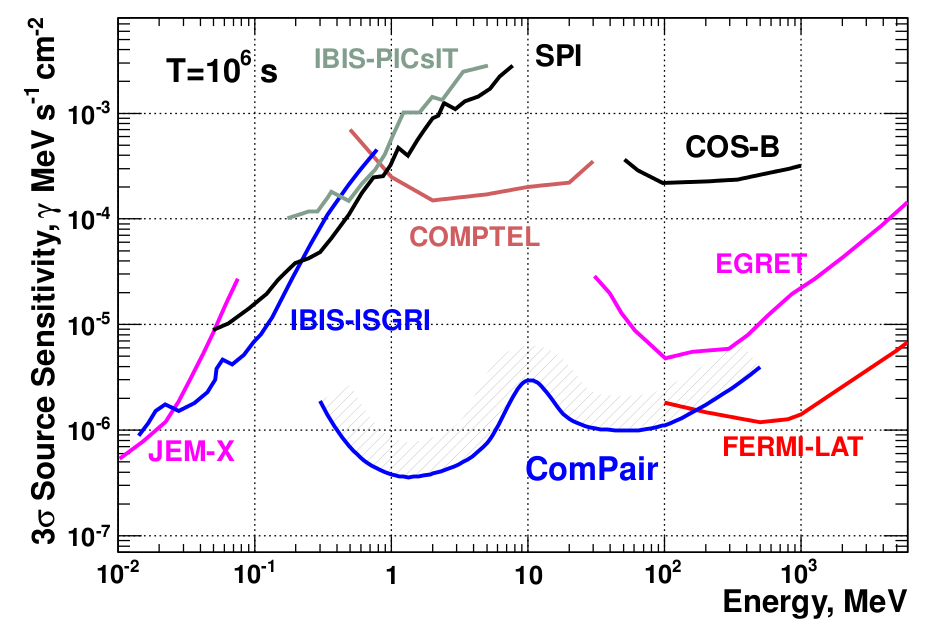}
\caption{Differential point source sensitivity of past and current missions together with the sensitivity expected for the proposed mission ComPair, which aims to close the sensitivity gap in the MeV energy range \cite{McEnery2015}. The figure is reproduced from \cite{Moiseev1035}.}
\label{fig:mevgap}
\end{wrapfigure}

To date, no successor mission has been approved to observe the gamma-ray sky at a sensitivity comparable to the LAT after \emph{Fermi} stops operations (presumably towards the end of this decade). However, several missions are currently being proposed; they primarily aim to extend the LAT sensitivity towards lower energies, to close the sensitivity gap of current and past missions around $\approx$ 1 MeV (``the MeV gap''). The last mission observing the sky at these energies was COMPTEL, which finished operations in the year 2000.  At a photon energy of  $\approx$ 1 MeV Compton scatterings are more frequent than pair creation used at higher energies. The techniques of the proposed missions PANGU and ComPair are based on a silicon strip tracker \cite{Wu964,McEnery2015,Moiseev1035}. In comparison to the LAT, no converter foils are inserted between the tracker layers. The sensitivity expected for the ComPair is shown in fig. \ref{fig:mevgap}. The angular resolution is expected to be $\approx 4^{\circ}$ at 1 MeV.

Another approach to detect Compton scattering events is to use gas Time Projection Chambers. This technique was tested successfully on two balloon flights with the SMILE instrument \cite{Komura1019}. The advantage compared to silicon trackers is that a higher angular resolution of $\approx 1^{\circ}$ can potentially be achieved. It should also be mentioned, that two missions with the ability to detect gamma rays at $\approx$ 1 GeV are expected to begin operations within the next year: CALET \cite{Cannady995} and DAMPE \cite{Tykhonov1193}. However, the primary aim of these satellites is to measure the cosmic-ray electron spectrum precisely. Their gamma-ray capabilities have not been fully evaluated yet, but they will likely not be comparable to the ones of the LAT. Driven by the success of the \emph{Fermi} mission, hopefully a new dedicated gamma-ray satellites will become reality soon and keep our eyes open at the fascinating gamma-ray sky.

\acknowledgments

I would like to thank the organizers of the invitation to give a rapporteur talk at the 34$^{th}$ ICRC and to write this article. I also want to thank the LAT team, particularly those working behind the scenes on running, calibrating and improving the instrument. A special thank you goes to the people who worked on the ``Pass  8'' event reconstruction over the past years.

\end{document}